\documentclass[3p,times,twocolumn]{elsarticle}

\usepackage{ecrc}

\volume{00}

\firstpage{1}

\journalname{Nuclear and Particle Physics Proceedings}

\runauth{}

\jid{nppp}

\jnltitlelogo{Nuclear and Particle Physics Proceedings}

\usepackage{amssymb}

\usepackage{amsmath}

\usepackage[figuresright]{rotating}
\usepackage{ctable}
\usepackage{siunitx}
\usepackage{booktabs,dcolumn}

\begin{document}

\begin{frontmatter}



\dochead{}

\title{Spectrum of Yang-Mills theory in 3 and 4 dimensions}


\author{Marco Frasca}
\ead{marcofrasca@mclink.it}
\address{Via Erasmo Gattamelata, 3, 00176 Rome (Italy)}

\begin{abstract}
We solve exactly the Dyson-Schwinger equations for Yang-Mills theory in 3 and 4 dimensions. This permits us to obtain the exact correlation functions till order 2. In this way, the spectrum of the theory is straightforwardly obtained and comparison with lattice data can be accomplished. The results are in exceedingly good agreement with an error well below 1\%. This extends both to 3 and 4 dimensions and varying the degree of the gauge group. These results provide a strong support to the value of the lattice computations and show once again how precise can be theoretical computations in quantum field theory.
\end{abstract}


\end{frontmatter}


\section{Introduction}

Yang-Mills theory, and more generally quantum chromodynamics (QCD), is known to be difficult to approach theoretically. Most results we know come from perturbation theory in a small coupling regime or from computations done on powerful computer facilities, using extended lattices, where the theory is mapped in the Euclidean case. Results about the propagators and the spectrum have been successfully obtained just in the latter case in 4 dimensions \cite{Lucini:2004my,Chen:2005mg,Bogolubsky:2007ud,Cucchieri:2007md,Oliveira:2007px} and \cite{Teper:1998te,Lucini:2002wg,Bringoltz:2006zg,Caselle:2011fy,Caselle:2011mn,Athenodorou:2016ebg} in 3 dimensions . 

So far, theoretical good results on the spectrum have been obtained using dispersion relations (see \cite{Narison:2007spa} and refs. therein). In general, theoretical derivations, that are obtained straightforwardly from Yang-Mills theory, are lacking in 4 dimensions, in the low-energy limit. The situation appears rather better in 3 dimensions, where some analysis have been performed, reaching agreement with lattice computations for the string tension \cite{Karabali:1997wk,Karabali:1998yq,Nair:2002yg,Karabali:2009rg} and the spectrum \cite{Leigh:2005dg,Leigh:2006vg}. A successful approach was devised by Karabali, Kim and Nair with a set of matrix variables to work in this case to obtain a wavefunction results of the theory \cite{Karabali:1995ps,Karabali:1997wk,Karabali:1998yq}. Karabali, Kim and Nair approach permits to obtain the string tension and higher order corrections to it \cite{Karabali:2009rg}. For the spectrum, a proper wavefunction was postulated \cite{Leigh:2005dg,Leigh:2006vg}. The agreement with lattice data was impressive but also logarithmic confinement was obtained. Generalizations to 4 dimensions of this approach were unsuccessful so far. 

About the question of confinement, in 4 dimensions a key role is played by the running coupling. This has been extensively discussed in literature \cite{Nesterenko:1999np,Nesterenko:2001st,Nesterenko:2003xb,Baldicchi:2007ic,Baldicchi:2007zn,Bogolubsky:2009dc,Duarte:2016iko,Deur:2016tte,Deur:2016bwq}. While in 3 dimensions the theory is super-renormalizable and no renormalization occurs, in 4 dimensions, due to the running coupling, a linear potential arises by combining this with the potential in the Wilson loop \cite{Gonzalez:2011zc,Deur:2016bwq}. In 3 dimensions, confinement occurs but with a logarithmic potential \cite{Frasca:2016sky} as also seen in \cite{Leigh:2005dg,Leigh:2006vg}.

In this contribution, we will show how Yang-Mills theory can be solved exactly both in 3 and 4 dimensions, obtaining the exact correlation functions till order 2, providing the spectrum in agreement with lattice results well below 1\%. Confinement is straightforwardly proved in 3 dimensions. This grants also a theoretical validation of lattice computations, if ever needed.

We consider a set of massive non-linear waves exactly solving the classical Yang-Mills equations \cite{Frasca:2016sky,Frasca:2015yva} at the ground state of the theory. This can be written in the form
\begin{equation}
      A_\mu^a=\eta_\mu^a\chi(x)+O\left(1/Ng^2\right)
\end{equation}
being $\eta_\mu^a$ a set of constants. Then, in $d=3+1$ these are
\begin{equation}
      \chi(x)=\mu(2/Ng^2)^\frac{1}{4}{\rm sn}(p\cdot x+\phi,-1)
\end{equation}
being $\mu$ a constant with the dimensions of a mass and $\phi$ an arbitrary phase. These are exact solutions provided the following dispersion relation holds
\begin{equation}
      p^2=\mu^2\sqrt{Ng^2/2}. 
\end{equation}
This is so because the theory is massless but, if we would add a mass term, changes in both the dispersion relation and the solution must be considered as we will see below. For $d=2+1$ the solution is
\begin{equation}
     \chi(x)=a\sqrt{2^\frac{1}{2}Ng^2}{\rm sn}(p\cdot x+\phi,-1)
\end{equation}
being $a$ a dimensionless constant. Now the dispersion relation becomes 
\begin{equation}
     p^2=a^2N^2g^4/\sqrt{2}.
\end{equation}
These solutions solve the 1P Dyson-Schwinger equation in the Landau gauge \cite{Frasca:2015yva}
\begin{flalign}
      &\partial^2G_{1\nu}^{a}(x)+gf^{abc}(
		\partial^\mu G_{2\mu\nu}^{bc}(0)+\partial^\mu G_{1\mu}^{b}(x)G_{1\nu}^{c}(x)- &\\
		&\partial_\nu G_{2\mu}^{\nu bc}(0)-\partial_\nu G_{1\mu}^{b}(x)G_{1}^{\mu c}(x)) \nonumber &\\
		&+gf^{abc}\partial^\mu G_{2\mu\nu}^{bc}(0)+gf^{abc}\partial^\mu(G_{1\mu}^{b}(x)G_{1\nu}^{c}(x)) \nonumber &\\
		&+g^2f^{abc}f^{cde}(G_{3\mu\nu}^{\mu bde}(0,0)
		+G_{2\mu\nu}^{bd}(0)G_{1}^{\mu e}(x) \nonumber &\\
	&+G_{2\nu\rho}^{eb}(0)G_{1}^{\rho d}(x)
	+G_{2\mu\nu}^{de}(0)G_{1}^{\mu b}(x)+ \nonumber &\\
	&G_{1}^{\mu b}(x)G_{1\mu}^{d}(x)G_{1\nu}^{e}(x))
		=0. \nonumber
\end{flalign}
This equation can be simplified to
\begin{flalign}
      &\partial^2G_{1\nu}^{a}(x)+g^2f^{abc}f^{cde}(G_{3\mu\nu}^{\mu bde}(0,0)
		+G_{2\mu\nu}^{bd}(0)G_{1}^{\mu e}(x) &\\
		&+G_{2\nu\rho}^{eb}(0)G_{1}^{\rho d}(x)
	+G_{2\mu\nu}^{de}(0)G_{1}^{\mu b}(x) \nonumber &\\
	&+G_{1}^{\mu b}(x)G_{1\mu}^{d}(x)G_{1\nu}^{e}(x))
		=0. \nonumber
\end{flalign}
with a ``mass term'' determined by the 2P function $G_2(0)$. This is a characteristic of the Dyson-Schwinger equations where lower order equations depend on solutions to higher order equations. In our case, this is not a concern. In $d=2+1$, this mass term can be set to zero because no renormalization of mass occurs \cite{Frasca:2016sky}. Indeed, the overall effect is to give a correction to the arbitrary dimensionless parameter $a$ and so, it is pointless. In $d=3+1$, this term should be accounted for as renormalization of mass occurs. The Dyson-Schwinger equation for the 2P function is in the Landau gauge can be written down as \cite{Frasca:2015yva} 
\begin{flalign}
      &\partial^2G_{2\nu\kappa}^{am}(x-y)+gf^{abc}(
		\partial^\mu G_{3\mu\nu\kappa}^{bcm}(0,x-y) &\\
		&+\partial^\mu G_{2\mu\kappa}^{bm}(x-y)G_{1\nu}^{c}(x)
		+\partial^\mu G_{1\mu}^{b}(x)G_{2\nu\kappa}^{cm}(x-y) \nonumber &\\
		&-\partial_\nu G_{3\mu\kappa}^{\mu bcm}(0,x-y)-\partial_\nu G_{2\mu\kappa}^{bm}(x-y)G_{1}^{\mu c}(x) \nonumber &\\
		&-\partial_\nu G_{1\mu}^{b}(x)G_{2\kappa}^{\mu cm}(x-y))
		+gf^{abc}\partial^\mu G_{3\mu\nu\kappa}^{bcm}(0,x-y) \nonumber &\\
		&+gf^{abc}\partial^\mu(G_{2\mu\kappa}^{bm}(x-y)G_{1\nu}^{c}(x))
				+gf^{abc}\partial^\mu(G_{1\mu}^{b}(x)G_{2\nu\kappa}^{cm}(x-y)) \nonumber &\\
		&+g^2f^{abc}f^{cde}(G_{4\mu\nu\kappa}^{\mu bdem}(0,0,x-y)
		+G_{3\mu\nu\kappa}^{bdm}(0,x-y)G_{1}^{\mu e}(x) \nonumber &\\
		&+G_{2\mu\nu}^{bd}(0)G_{2\kappa}^{\mu em}(x-y)
	+G_{3\nu\rho\kappa}^{acm}(0,x-y)G_{1}^{\rho b}(x)  \nonumber &\\
	&+G_{2\nu\rho}^{eb}(0)G_{2\kappa}^{\rho dm}(x-y)
	+G_{2\nu\rho}^{de}(0)G_{2\kappa}^{\rho bm}(x-y) \nonumber &\\
	&+G_{1}^{\mu b}(x)G_{3\mu\nu\kappa}^{dem}(0,x-y)+
	G_{2\kappa}^{\mu bm}(x-y)G_{1\mu}^{d}(x)G_{1\nu}^{e}(x)+  \nonumber &\\
	&G_{1}^{\mu b}(x)G_{2\mu\kappa}^{dm}(x-y)G_{1\nu}^{e}(x)+
	G_{1}^{\mu b}(x)G_{1\mu}^{d}(x)G_{2\nu\kappa}^{em}(x-y)) \nonumber &\\
	&=gf^{abc}(\partial_\nu K^{bcm}_{3\kappa}(0,x-y)+\partial_\nu (\bar P^{b}_1(x)K^{cm}_{2\kappa}(x-y))) \nonumber &\\
	&+\partial_\nu (\bar K^{bm}_{2\kappa}(x-y)P^{c}_1(x)))
	+\delta_{am}g_{\nu\kappa}\delta^4(x-y). \nonumber
\end{flalign}
In the Landau gauge, these equations take a very simple form because the 2P function can be cast in the form
\begin{equation}
      G_{2\nu\kappa}^{am}(p)=\delta_{am}\left(g_{\nu\kappa}-\frac{p_\nu p_\kappa}{p^2}\right)\Delta(p).
\end{equation}
Indeed, these Dyson-Schwinger equations can be rewritten in the following form \cite{Frasca:2017mrh}. The 1P function is given by the equation
\begin{equation}
      \partial^2\chi(x)+2Ng^2\delta\mu^2\chi(x)+Ng^2\chi^3(x)=0,
\end{equation}
obtaining, in four dimensions, the mass correction
\begin{equation}
	\delta\mu^2=\int\frac{d^4p}{(2\pi)^4}\Delta(p),
\end{equation}
and, for the 2P function, one has
\begin{equation}
	\partial^2\Delta(x-y)+2Ng^2\delta\mu^2\Delta(x-y)+3Ng^2\phi^2(x)\Delta(x-y)=\delta^4(x-y).
\end{equation}

\section{Case $d=2+1$}

We start by discussing the 3 dimensional case. The equation for the propagator can be solved exactly, given the 1P function, and the solution is the following \cite{Frasca:2016sky}
\begin{equation}
  \Delta(p)=\sum_{n=0}^\infty\frac{B_n}{p^2-m_n^2+i\epsilon}
\end{equation}
being
\begin{equation}
	B_n=(2n+1)^2\frac{\pi^3}{4K^3(-1)}\frac{e^{-(n+\frac{1}{2})\pi}}{1+e^{-(2n+1)\pi}}
\end{equation}
and the mass spectrum
\begin{equation}
	m_n=(2n+1)\frac{\pi}{2K(-1)}a\frac{Ng^2}{2^\frac{1}{4}}
\end{equation}
being $K(-1)$ the complete elliptic integral of the first kind. We note the presence of the dimensionless $a$ factor. The theory is super-renormalizable in 3 dimensions so, the $\delta\mu^2$ term is harmless here and can be absorbed into the arbitrary $a$ factor. Due to the properties of the gluon propagator, the Wilson loop can be easily evaluated yielding the potential \cite{Frasca:2016sky}
\begin{equation}
    V_{YM}(r)=-\frac{g^2}{4\pi}C_2(R)\sum_{n=0}^\infty B_nK_0(m_nr)
\end{equation}
being $K_0$ a modified Bessel function of order 0. This is the potential given by Leigh\&al. \cite{Leigh:2005dg,Leigh:2006vg}. Yang-Mills theory in 2+1 dimensions is confining but the potential increases just logarithmically. This can be seen immediately by noticing that, at small distances, $K_0(m_nr)=-\ln(m_nr/2)+\ldots$. Then, the string tension can be evaluated to \cite{Frasca:2016sky}
\begin{equation}
\label{eq:sigma}
    \sigma^R_{KKN}=\sigma_{KKN}\frac{\pi}{2^\frac{5}{4}K(-1)}a
\end{equation}
being $\sigma_{KKN}$ the string tension given by Karabali-Kim-Nair \cite{Karabali:1997wk,Karabali:1998yq}. The power of the $a$ factor will prove to be crucial in the determination of the spectrum. This choice of the string tension grants the exceedingly good agreement with the lattice spectrum, as we will see in a moment. 

The spectrum can be straightforwardly evaluated and compared to the lattice results \cite{Athenodorou:2016ebg}. One has \cite{Frasca:2016sky}
\begin{equation}
     \frac{m_n}{\sqrt{\sigma}}=(2n+1)\frac{\pi}{2\cdot 2^\frac{1}{4}K(-1)}a\frac{Ng^2}{\sqrt{\sigma}}
\end{equation}
or
\begin{equation}
     \frac{m_n}{\sqrt{\sigma}}=(2n+1)\cdot 5.032050686\ldots\cdot\sqrt{a}\frac{1}{\sqrt{1-\frac{1}{N^2}}},
\end{equation}
where use has been made of eq.~(\ref{eq:sigma}) for the string tension. We see that the parameter $a$ enters with a square root. This can be obtained from preceding results in literature and is now completely justified by the theory. The proper value is $a=2/3$ \cite{Isgur:1984bm,Johnson:2000qz}. The comparison with lattice data for the ground state of the theory is given in Tab.~\ref{tab1}. The agreement is stunning and well-below 1\% for all $N$.
\ctable[
caption = {\label{tab1} Comparison for the ground state at varying $N$ and for $N\rightarrow\infty$.},
center
]
{@{}lllll@{}}{
		\tiny{(a)~Athenodorou\&Teper, JHEP 1702, 015 (2017).}
}{
    \toprule
 {$N$}       & \multicolumn{1}{l}{Lattice$^{(a)}$}    & \multicolumn{1}{l}{Theoretical} & \multicolumn{1}{l}{Error} \\
    \midrule    
  2        & 4.7367(55) & 4.744262871 & 0.16\% \\
  3        & 4.3683(73) & 4.357883714 & 0.2\% \\
  4        & 4.242(9)   & 4.243397712 & 0.03\% \\
 $\infty$ & 4.116(6)   & 4.108652166 & 0.18\% \\
    \bottomrule
}

\section{Case $d=3+1$}

In four dimensions, we cannot neglect the mass correction $\delta\mu^2$ anymore and so, the classical solutions to consider must be changed in the following way
\begin{flalign}
&\chi(x)=\sqrt{\frac{2\mu^4}{m^2+\sqrt{m^4+2\lambda\mu^4}}}\times &\\
		&{\rm sn}\left(p\cdot x+\phi,
		\frac{m^2-\sqrt{m^4+2Ng^2\mu^4}}{m^2+\sqrt{m^4+2Ng^2\mu^4}}\right) \nonumber
\end{flalign}
with the dispersion relation
\begin{equation}
    p^2=m^2+\frac{Ng^2\mu^4}{m^2+\sqrt{m^4+2Ng^2\mu^4}}.
\end{equation}
The mass correction is given by
\begin{equation}
   m^2=2Ng^2G_2(0).
\end{equation}
As we will see, this mass correction has a negative sign lowering the bare mass spectrum and fitting properly the lattice data. Due to the correction arising from the renormalization of mass, the propagator changes accordingly. To work it out, we assume that $G_2(0)$ should be evaluated iteratively. This will permit us to avoid numerical computations obtaining anyway a consistent closed form solution. Indeed, one has \cite{Frasca:2017mrh}
\begin{flalign}
&\Delta(p)=\sqrt{m^2+\mu^2\sqrt{Ng^2/2}}Z_\Delta(m^2,Ng^2)\frac{2\pi^3}{K^3(k^2(m))}\times &\\
&\sum_{n=0}^\infty (-1)^n(2n+1)^2\frac{q^{n+1/2}}{1-q^{2n+1}}\frac{1}{p^2-m_n^2+i\epsilon}.  \nonumber
\end{flalign}
We have set
\begin{equation} 
k^2(m)=\frac{m^2-\sqrt{m^4+2Ng^2\mu^4}}{m^2+\sqrt{m^4+2Ng^2\mu^4}},
\end{equation}
being
\begin{equation}
\lim_{m\rightarrow 0}\sqrt{m^2+\mu^2\sqrt{Ng^2/2}}Z_\Delta(m^2,Ng^2)=\frac{1}{8}
\end{equation}
and $q=\exp\left(-\pi K'(k^2)/K(k^2)\right)$ with $K'(k^2)=K(1+k^2)$. There are two key equations here. The mass spectrum
\begin{equation}
m_n(m)=(2n+1)\frac{\pi}{2K(k^2)}\sqrt{m^2+\frac{Ng^2\mu^4}{m^2+\sqrt{m^4+2Ng^2\mu^4}}},
\end{equation}
with $k^2$ depending on $m$, and the mass shift $m^2=2Ng^2\Delta(0)$. To evaluate the mass spectrum, we need to solve the following gap equation
\begin{flalign}
&m^2=2Ng^2\sqrt{m^2+\mu^2\sqrt{Ng^2/2}}Z_\Delta(m^2,Ng^2)\times &\\
&\int\frac{d^4p}{(2\pi)^4}\sum_{n=0}^\infty (-1)^n(2n+1)^2\frac{2\pi^3}{K^3(k^2(m))}\times \nonumber &\\
&\frac{q^{n+1/2}}{1-q^{2n+1}}\frac{1}{p^2-m_n^2+i\epsilon}. \nonumber
\end{flalign}
This represents a gap equation for the theory that should be solved to get the spectrum. The best analytical approach is to assume the shift small and solve it iteratively. We will see that, in this way, a consistent solution is obtained. So, we take for the first iteration $m=0$ giving
\begin{flalign}
&m^2\approx 2Ng^2\int\frac{d^4p}{(2\pi)^4}\sum_{n=0}^\infty(2n+1)^2\frac{\pi^3}{4K^3(-1)}\times &\\
&\frac{e^{-(n+\frac{1}{2})\pi}}{1+e^{-(2n+1)\pi}}
	\frac{1}{p^2-m_n^2(0)+i0}. \nonumber
\end{flalign}
The evaluation of the integral is standard and can be accomplished by using dimensional regularization. This yields ($\epsilon=4-d$)
\begin{flalign}
&m^2(\epsilon)=Ng^2\sum_{n=0}^\infty\left[-\frac{2}{\epsilon}+(\gamma-1)+\ln\frac{m_n^2(0)}{4\pi\Lambda^2}+O(\epsilon)\right]\times &\\
&(2n+1)^2\frac{\pi}{32K^3(-1)}\frac{e^{-(n+\frac{1}{2})\pi}}{1+e^{-(2n+1)\pi}}m^2_n(0). \nonumber
\end{flalign}
Here we see the appearance of the cut-off $\Lambda$ arising from dimensional regularization. From this we extract the finite part to use in our computations
\begin{flalign}
&m^2_{finite}=Ng^2(\gamma-1)\sum_{n=0}^\infty(2n+1)^2\frac{\pi}{32K^3(-1)}\times &\\
&\frac{e^{-(n+\frac{1}{2})\pi}}{1+e^{-(2n+1)\pi}}m^2_n(0). \nonumber
\end{flalign}
A straightforward computation gives for the finite part
\begin{equation}
    m^2_{finite}={\bar m}^2Ng^2\sigma
\end{equation}
being $\sigma=\sqrt{Ng^2/2}\mu^2$, the string tension to use to compare with lattice results, and ${\bar m}^2=-0.03212775693\ldots$ that is negative, shifting the spectrum downward. The final formula, to use for comparison with lattice data given in \cite{Lucini:2004my}, is
\begin{flalign}
\label{eq:ms}
 &\frac{m_n}{\sqrt{\sigma}}\approx(2n+1)\frac{\pi}{2K(-1)}\times &\\
 &		\left[1+\left(\frac{1}{4}-
		\frac{1}{2}\frac{K(\sqrt{2}/2)-E(\sqrt{2}/2)}{K(\sqrt{2}/2)}
		\right){\bar m}^2\frac{2N^2}{\beta}\right]. \nonumber
\end{flalign}
Here $K(k)$ is the complete elliptic integral of the first kind and $E(k)$ is of the second kind. We have introduced $\beta=2N/g^2$ in place of $Ng^2$ to make easier the comparison with lattice data that is given in Tab.~\ref{tab2} for the would-be ground state. Indeed, we point out that we have put $n=1$ into eq.(\ref{eq:ms}) as lattice data do not seem to display the $n=0$ case while this clearly appears for the first excited state as explained below and is also clearly seen in $d=2+1$. The agreement is exceedingly good hinting to some possible problems in the lattice computations for SU(2) and SU(6) but those data are quite old by now with respect to the 2+1 case.
\ctable[
caption = {\label{tab2} Comparison for the ground state at varying $N$ and fixing $\beta$ to the proper lattice value.},
center
]
{@{}lllll@{}}{
  \tiny{(a)~Lucini, Teper, Wenger, JHEP 0406, 012 (2004).}
}{
\toprule
{$N$}      & \multicolumn{1}{l}{Lattice$^{(a)}$}    & \multicolumn{1}{l}{Theoretical}    & \multicolumn{1}{l}{$\beta^{(a)}$}     & \multicolumn{1}{l}{Error} \\
\midrule
2        & 3.78(7)    & 3.550927197    & 2.4265      & 6.0\% \\
3        & 3.55(7)    & 3.555252334    & 6.0625      & 0.1\% \\
4        & 3.56(11)   & 3.556337890    & 11.085      & 0.1\% \\
6        & 3.25(9)    & 3.557102106    & 25.452      & 8.6\% \\
8        & 3.55(12)   & 3.557471208    & 45.70       & 0.2\% \\
\bottomrule
}

In order to compute the first excited state, we just note that, in \cite{Lucini:2004my}, states with $m_n=(2n+1)m_0$, $m_{n_1,n_2}=(2n_1+2n_2+2)m_0$ and so on are obtained, being $m_0$ the ground state we showed above. So, the lowest excited state is obtained by $m_{0,0}=2m_0$ that yields Tab.~\ref{tab3}.
\ctable[
caption = {\label{tab3} Comparison for the first excited state at varying $N$ and fixing $\beta$ to the proper lattice value.},
center
]
{@{}lllll@{}}{
  \tiny{(a)~Lucini, Teper, Wenger, JHEP 0406, 012 (2004).}
}{
\toprule
{$N$}      & \multicolumn{1}{l}{Lattice$^{(a)}$}    & \multicolumn{1}{l}{Theoretical}    & \multicolumn{1}{l}{$\beta^{(a)}$}     & \multicolumn{1}{l}{Error} \\
\midrule
2        & 5.45(11)   & 4.734569596    & 2.4265      & 13.0\% \\ 
3        & 4.78(9)    & 4.740336446    & 6.0625      & 0.8\% \\
4        & 4.85(16)   & 4.741783854    & 11.085      & 2.0\% \\
6        & 4.73(15)   & 4.742802808    & 25.452      & 0.3\% \\
8        & 4.73(22)   & 4.743294944    & 45.70       & 0.3\% \\
\bottomrule
}
Agreement is exceedingly good as in the previous cases. Problems on lattice computations seem to be evident for SU(2). Anyway, for SU(3) the agreement is excellent and also the tendency of the spectrum, at increasing $N$, seems well recovered. We emphasize that we have used the values of $\beta$ given in \cite{Lucini:2004my} in our formula (\ref{eq:ms}) to reach this agreement.

\section{Conclusions}

We were able to solve exactly the Dyson-Schwinger equations for the Yang-Mills theory obtaining the correlation functions till order 2, both in 3 and 4 dimensions. This permitted us to obtain the spectrum of the theory to compare with lattice data. Lattice data are quite recent for 3 dimensions and we get a stunning agreement (below 1\%) with our theoretical computation for the ground state of the theory, also varying the degree of the SU(N) group. In 4 dimensions, lattice data are quite older but, anyway, the agreement we obtained is at the same level (below 1\%) and, also in this case, at varying the degree of the gauge group. Some exceptions occur, e.g. SU(2) for the ground and the excited state as also the missing $n=0$ state, but this can be ascribed to the lattice data, mostly if we expect a similar behavior of the Yang-Mills theory in 3 and 4 dimensions. At this stage, known also the availability of largely improved computer facilities with respect to 2004, it is urged to see fresher lattice data for this case as done for the lower dimensional case. It is important to notice that this theoretical results give a strong support to lattice computations and show again how powerful and precise quantum field theory can be.







\end{document}